\documentclass[10pt,aps,prd,showpacs,footinbib,twocolumn]{revtex4-1}
\usepackage[usenames,dvipsnames]{color}
\usepackage{amsfonts}
\usepackage{amssymb}
\usepackage{amsmath}
\usepackage{graphicx}
\usepackage{epstopdf}
\usepackage[colorlinks]{hyperref}
\usepackage[normalem]{ulem}

\begin{document}

\title{Scalar charges and pulsar-timing observables in the presence of nonminimally coupled scalar fields}

\author{Raissa F.\ P.\ Mendes}
\email{rfpmendes@id.uff.br}
\affiliation{Instituto de F\'isica, Universidade Federal Fluminense, Niter\'oi, 
Rio de Janeiro, 24210-346, Brazil.}
\author{Tulio Ottoni}
\email{tulioottoni@id.uff.br}
\affiliation{Instituto de F\'isica, Universidade Federal Fluminense, Niter\'oi, 
Rio de Janeiro, 24210-346, Brazil.}

\date{\today}

\begin{abstract}
Pulsar-timing has become a celebrated tool for probing modifications to general relativity in the strong-field surroundings of neutron stars. Here we investigate whether scalar-tensor theories that incorporate a nonminimally coupled scalar degree of freedom may pass pulsar-timing tests, by computing the scalar charges entering such observables. In particular we show that for positive values of the nonminimal coupling $\xi$, pulsar-timing constraints may be evaded even in the presence of spontaneous scalarization.
\end{abstract}

\pacs{04.50.Kd, 04.40.Dg, 04.80.Cc, 97.60.Gb} 

\maketitle

\section{Introduction}

Perhaps the simplest way in which general relativity (GR) can be modified is through the coupling to a new scalar degree of freedom $\Phi$, as accomplished by the general scalar-tensor action
\begin{equation}\label{eq:actionF}
S_g \! = \! \frac{1}{16\pi G} \int \!\! d^4 x \sqrt{-g} \left[ F(\Phi) R - \! Z(\Phi) \nabla_\mu \Phi \nabla^\mu \Phi - V(\Phi) \right].
\end{equation}
This class of theories naturally implements Dirac's idea of varying fundamental ``constants'' \cite{Dirac1937,Uzan2003}, as $G_\textrm{eff} \equiv G/F(\Phi)$ can be interpreted as a time- and space-dependent effective gravitational coupling. A well-motivated form for $F(\Phi)$ is the standard nonminimal coupling (SNMC),
\begin{equation} \label{eq:Fxi}
F(\Phi) = 1 - 8\pi \xi \Phi^2,
\end{equation}
where $\xi \in \mathbb{R}$. Its motivation ranges from fundamental considerations---such as those arising from the quantization of the classical theory in a curved spacetime \cite{Birrell1984}---to its usefulness in cosmological model-building, specially in inflationary scenarios \cite{Faraoni1996,Faraoni2001}. 

In the class of scalar-tensor theories (STTs) defined by Eq.~(\ref{eq:actionF}), post-Newtonian (PN) deviations from GR are proportional to $(dF/d\Phi)|_{\Phi_0}$, where $\Phi_0$ is the scalar field value at the current cosmological epoch \cite{Damour1992}. For the SNMC, $(dF/d\Phi)|_{\Phi_0} \propto \Phi_0$, and so the observed agreement \cite{Will2006} between solar system observations and GR's predictions implies that $\Phi_0$ must be close to zero---but does not limit the viable range of $\xi$.
Interestingly, however, even if $\Phi_0 = 0$, STTs may still differ considerably from GR in their predictions for neutron stars (NS), due to a nonperturbative, strong-field effect known as spontaneous scalarization \cite{Damour1993}. This effect, which has long been known to happen for sufficiently negative values of the nonminimal coupling $\xi$, is characterized by the formation of a scalar cloud that modifies the star's equilibrium and perturbative properties \cite{Salgado1998,AltahaMotahar2017,Sotani2004,Sotani2005,Mendes2018}, and has dramatic implications, most notably for pulsar-timing observables \cite{Will1993,Damour1996,Wex2014}. 
Indeed, the inconsistency between pulsar-timing data and certain aspects of NS phenomenology in STTs, such as the existence of scalar dipole radiation, of scalar-field induced variations in the NS moment of inertia, and so on, can be used to rule out almost the entire range $\xi \lesssim - 2.2$ of field couplings allowing for spontaneous scalarization (in the case where $V(\Phi) = 0$, which we will refer to as ``massless'' for simplicity).

More recently, it has been shown that positive values of the nonminimal coupling $\xi$ can give rise to a similar spontaneous scalarization effect around sufficiently compact neutron stars, i.e., stars with $G M/(R c^2) \gtrsim 0.26$, where $G$ is Newton's constant, $c$ is the speed of light, and $M$ and $R$ denote the NS mass and radius \cite{Mendes2015,Mendes2016,Podkowka2018}. Massless STTs with $\xi>0$ (which include the conformal coupling $\xi = 1/6$ as a particular case) are known to provide consistent cosmological scenarios \cite{Damour1993b,Esposito-Farese2001,Anderson2016,Anderson2017}, but remain largely unconstrained by astrophysical observations. 

The purpose of the present work is to explore how the main post-Keplerian pulsar-timing observables---the Einstein time delay $\gamma$, the rate of periastron advance $\dot{\omega}$, and the rate of decay of the orbital period $\dot{P}_b$---are modified around scalarized NSs in STTs, and investigate whether pulsar-timing data could also be used to constrain massless STTs with $\xi >0$. Interestingly, we find some crucial differences between the nature of spontaneous scalarization in the $\xi <0$ and $\xi >0$ cases, which reduces the effectiveness of pulsar timing observations in placing new constraints, even in the presence of spontaneous scalarization. In particular, the main scalar charge, $\alpha_A$, entering these observables is typically suppressed in the $\xi >0$ case, and becomes progressively smaller as $\xi$ increases. As a consequence, even though the presence of dipolar scalar radiation in STTs gives a contribution to $\dot{P}_b$ which is enhanced by a factor of $(c/v)^2$ (where $v$ is the relative orbital velocity) with respect to GR, this term is suppressed by the smallness of the scalar charge, and $\dot{P}_b$ becomes dominated by the usual quadrupolar contribution. 
By exploring the dependence of $\alpha_A$ on the local scalar field environment, we also argue that the feedback mechanism responsible for the effect of \textit{dynamical} scalarization found in some STTs \cite{Barausse2013,Palenzuela2014,Sampson2014,Taniguchi2015, Sennett2016} will likely be absent when $\xi>0$.

This work is organized as follows. In Sec.~\ref{sec:framework} we discuss in more detail the framework we consider, including our choices of NS equations of state. In Sec.~\ref{sec:observables} we discuss how the pulsar-timing observables $\{\gamma, \dot{\omega}, \dot{P}_b\}$ are modified in STTs and briefly review how to compute the scalar charges $\{\alpha_A, \beta_A, k_A\}$ that enter such observables. Our main results are presented in Sec.~\ref{sec:results} and final considerations are made in Sec.~\ref{sec:conclusion}. A toy model presented in Appendix \ref{sec:toymodel} aims to elucidate in a simpler setting some of the features described in Sec.~\ref{sec:results}. We adopt natural units in which $c = G = 1$ unless specified.

\section{Framework} \label{sec:framework}

\subsection{Field equations}

Neutron stars can be studied in STTs by adding the contribution from the stellar fluid to the gravitational action (\ref{eq:actionF}):
\begin{equation} \label{eq:Sgm}
S = S_g + S_m [\Xi_m; g_{\mu\nu}].
\end{equation}
We assume that the energy-momentum tensor of the matter fields $\Xi_m$, $T^{\mu\nu} \equiv (2/\sqrt{-g}) \delta S_m/\delta g_{\mu\nu}$, has the form of a perfect fluid:
\begin{equation}\label{eq:Tmunu}
T^{\mu\nu} = (\epsilon + p) u^\mu u^\nu + p g^{\mu\nu},
\end{equation}
where $p$ and $\epsilon$ denote the pressure and energy density in the fluid's rest frame and $u^\mu$ is the four-velocity of fluid elements. 

For definiteness, in this work we will focus on massless scalar fields with no self-coupling, setting $V(\Phi) = 0$ in Eq.~(\ref{eq:actionF}). A mass term, for instance, would have the effect of delaying the onset of spontaneous scalarization to larger values of $|\xi|$  \cite{Ramazanoglu2016}, while other choices of $V(\Phi)$ could lead to a different phenomenology, including screening mechanisms \cite{Brax2012} (see also Refs.~\cite{Zhang2017,Zhang2019} in the context of pulsar-timing).
An additional simplification to Eq.~(\ref{eq:actionF}) results from exploiting its invariance under a scalar field redefinition, $\Phi \to \varphi(\Phi)$, to set $Z(\Phi)$ to a constant. Then, the only free parameter is the nonminimal coupling constant $\xi$, once $F(\Phi)$ is chosen as in Eq.~(\ref{eq:Fxi}).

For numerical calculations, it is often convenient to define the conformally rescaled (Einstein-frame) metric
\begin{equation} \label{eq:EFmetric}
g_{\mu\nu}^* \equiv F(\Phi) g_{\mu\nu}
\end{equation}
and redefine the scalar field, $\Phi \to \varphi(\Phi)$, so that 
\begin{equation} \label{eq:phiredef}
\left( \frac{d\varphi}{d\Phi} \right)^2 = \frac{3}{4 F(\Phi)^2} \left( \frac{d F(\Phi)}{d\Phi}\right)^2 + \frac{Z(\Phi)}{2 F(\Phi)}.
\end{equation}
These transformations turn Eq.~(\ref{eq:Sgm}) into 
\begin{align}
S & = \frac{1}{16 \pi} \int d^4 x \sqrt{-g^*} \left( R^* - 2 g^{\mu\nu}_* \partial_\mu \varphi \partial_\nu \varphi \right) \nonumber \\
&+ S_m[\Xi_m; g^*_{\mu\nu}/F(\Phi(\varphi))],
\end{align}
and therefore re-express the coupling between scalar field and geometry as a coupling to matter.
The field equations in the Einstein frame take a simpler form,
\begin{align}
  & G_{\mu\nu}^* - 2 \partial_\mu \varphi \partial_\nu \varphi + g^*_{\mu\nu} g^{\sigma\rho}_* \partial_\sigma \varphi  \partial_\rho \varphi  = 8\pi T_{\mu\nu} F(\varphi)^{-1} ,\label{eq:metric_eq}
	\\
  &	\nabla_*^\mu \nabla^*_\mu \varphi = - 4 \pi F(\varphi)^{-2} \alpha(\varphi) T,
	\label{eq:phi_eq}
\end{align} 
where $T \equiv g_{\mu\nu} T^{\mu\nu} = 3 p - \epsilon$ is the trace of the energy-momentum tensor (\ref{eq:Tmunu}),
\begin{equation}
\alpha(\varphi) \equiv - \frac{1}{2} \frac{d \ln F(\Phi(\varphi))}{d\varphi},
\end{equation}
and all quantities marked with an asterisk are computed from the metric (\ref{eq:EFmetric}). We emphasize that although we restore to practical computations in the Einstein frame, our results can be easily translated to the Jordan frame and our physical conclusions are frame-independent.

For $F(\Phi)$ given by Eq.~(\ref{eq:Fxi}), and adopting a canonical normalization, $Z(\Phi) = 8\pi$, the field redefinition (\ref{eq:phiredef}) gives
\begin{equation} \label{eq:phiredef2}
\frac{d\varphi}{d\Phi} = 2 \sqrt{\pi} \frac{\sqrt{1-8\pi \xi (1-6\xi) \Phi^2}}{1-8\pi \xi \Phi^2},
\end{equation}
from which we can see that for $\xi >0$ the finite domain $\Phi \in (-\Phi_\textrm{cr},\Phi_\textrm{cr})$, with $\Phi_\textrm{cr}\equiv 1/\sqrt{8\pi \xi}$, is mapped into $\varphi \in (-\infty, \infty)$. No such restriction arises for $\xi <0$. If one interprets $G_\textrm{eff} = G/F(\Phi)$ as an effective gravitational coupling, gravity becomes weaker in the presence of the scalar field when $\xi <0$, and stronger if $\xi >0$, becoming infinitely attractive as $\Phi \to \pm \Phi_\textrm{cr}$.

\begin{figure}[t]
\includegraphics[width=\linewidth]{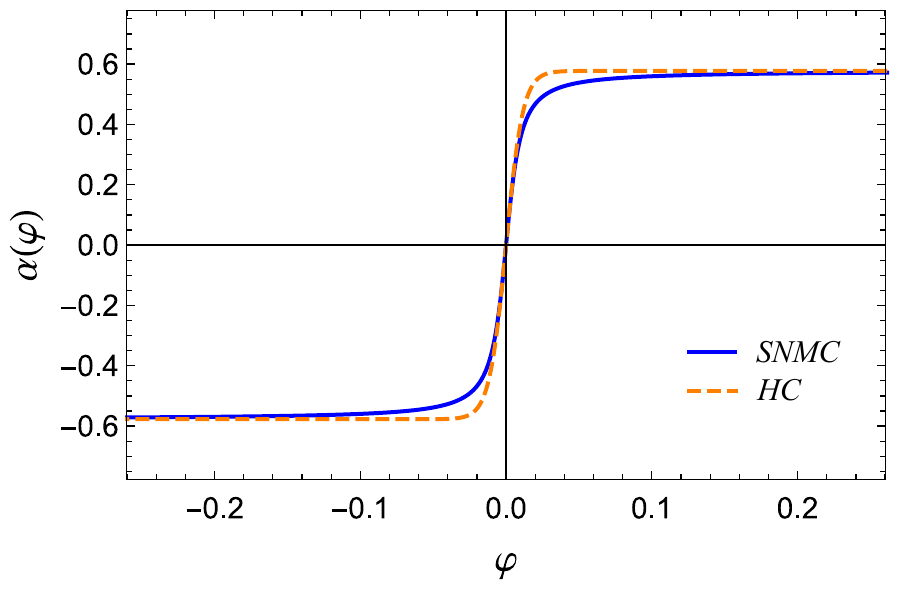}
\caption{Effective coupling $\alpha(\varphi)$ in terms of the Einstein-frame scalar field for the SNMC (solid blue) and the HC model of Eq.~(\ref{eq:HC2}) (dashed orange), both for $\xi = 25$. The HC model qualitatively reproduces the overall features of the SNMC, with the quantitative agreement improving as $|\xi|$ increases.}
\label{fig:NMCvsMO}
\end{figure}

Although Eq.~(\ref{eq:phiredef2}) can be integrated in terms of elementary functions, the inverse transformation $\Phi(\varphi)$ must be obtained numerically. Therefore, we will find it useful in this work to consider not only $F(\Phi)$ as given by Eq.~(\ref{eq:Fxi}) but also the following hyperbolic coupling (HC) model, already expressed in terms of the Einstein-frame field:
\begin{equation} \label{eq:HC1}
F(\varphi) = \left[ \cosh \left(2 \sqrt{3} \xi \varphi \right) \right]^{-1/(3\xi)},
\end{equation}
so that
\begin{equation} \label{eq:HC2}
\alpha(\varphi) = \frac{1}{\sqrt{3}} \tanh \left( 2\sqrt{3} \xi \varphi \right).
\end{equation}
This coupling function was suggested in Ref.~\cite{Mendes2016} (with $\beta = 2 \xi$) as a useful analytical approximation to a standard nonminimally coupled scalar field (see Fig.~\ref{fig:NMCvsMO}), and has been considered in Refs.~\cite{Anderson2019,Anderson2019a} in the context of pulsar-timing observations. Note that the SNMC and HC agree up to a cubic term in an expansion around $\varphi = 0$: In both cases $F(\varphi) = 1 - 2 \xi \varphi^2 + O(\varphi^4)$.

As mentioned in the introduction, solar system observations constrain the scalar field value at the current cosmological epoch, which we denote as $\Phi_0$ (with $\varphi_0 \equiv \varphi(\Phi_0)$ in the Einstein frame), to be very small. For instance, the parametrized post-Newtonian parameter $\gamma_\textrm{PPN}$ \cite{Will2006}, which takes the form
\begin{equation}
1- \gamma_\textrm{PPN} = \frac{(dF /d\Phi)^2}{ZF + 2 (dF/d\Phi)^2}
\end{equation}
for the theories described by Eq.~(\ref{eq:actionF}), 
is subject to the Cassini bound  $|1 - \gamma_\textrm{PPN}| \lesssim 2.1 \times 10^{-5}$ \cite{Bertotti2003}. The bound translates into $|\Phi_0| \lesssim 4.6 \times 10^{-4} /|\xi| $, which becomes more stringent as $|\xi|$ increases. In what follows we will typically fix the asymptotic, cosmological value of the scalar field to be $\Phi_0 = 0$ (or $\varphi_0 = 0$), but will also discuss some features of the $\Phi_0 \neq 0$ case.


\subsection{Equation of state}

In this work we will adopt three theoretical equations of state (EoS) for nuclear matter, the SLy \cite{Douchin2001}, ENG \cite{Engvik1995}, and MPA1 \cite{Muther1987} models. In GR, the sequence of equilibrium configurations generated by these EoS is causal (in the sense that the speed of sound does not exceed the speed of light inside any stable star) and has a large enough maximum mass to accommodate the observation of a $\sim 2 M_\odot$ NS \cite{Antoniadis2013}. Additionally, these EoS allow for stable NSs that are compact enough to trigger a spontaneous scalarization effect for both positive and negative values of the coupling constant $\xi$ (see Sec.~\ref{sec:scalarization} below). 

Instead of implementing these EoS through the interpolation of tabulated points, we shall approximate them by piecewise polytropes, adopting the parametrization developed in Ref.~\cite{Read2009}. This parametrization was shown to accurately reproduce the main NS properties predicted by theoretical EoS \cite{Read2009}, as well as NS scalar charges in STTs \cite{Anderson2019}. 


\subsection{Spontaneous scalarization windows} \label{sec:scalarization}

One of the most interesting phenomenological aspects of STTs in astrophysical scenarios is the spontaneous scalarization effect \cite{Damour1993}. 
In this section we briefly review the basic ideas behind this effect, and present the regions in parameter space where it takes place for the EoS employed in this work.

A careful inspection of the field equations (\ref{eq:metric_eq}) and (\ref{eq:phi_eq}) readily shows that a trivial scalar field profile, $\varphi = 0$, together with general-relativistic metric and matter configurations, form a solution of the field equations. However, for some relativistic stars, this trivial, GR-like solution may not be stable under scalar field perturbations \cite{Harada1997}. This can be seen by expanding Eqs.~(\ref{eq:metric_eq}) and (\ref{eq:phi_eq}) around $\varphi = 0$. To linear order in the perturbed quantities, the metric and fluid variables are not modified, while for the scalar field perturbation one obtains
\begin{equation}\label{eq:perturb}
\Box^{(0)} \delta \varphi = - 8 \pi \xi T^{(0)} \delta \varphi,
\end{equation}
where the index $(0)$ labels background quantities.
In the right-hand side of Eq.~(\ref{eq:perturb}), the combination $m_\textrm{eff}^2 \equiv - 8\pi \xi T^{(0)}$ can be loosely interpreted as an effective mass squared, and the fact that this can be negative signals the possible appearance of (tachyonic-like) instabilities (see Refs.~\cite{Lima2010a,Lima2010,Mendes2014} for a quantum analogue). 
The nonlinear development of this instability is the scalarization phenomenon: the spontaneous development of a cloud of scalar field around the star \cite{Novak1998,Gerosa2016,Mendes2016}. 

\begin{figure}[t]
\includegraphics[width=\linewidth]{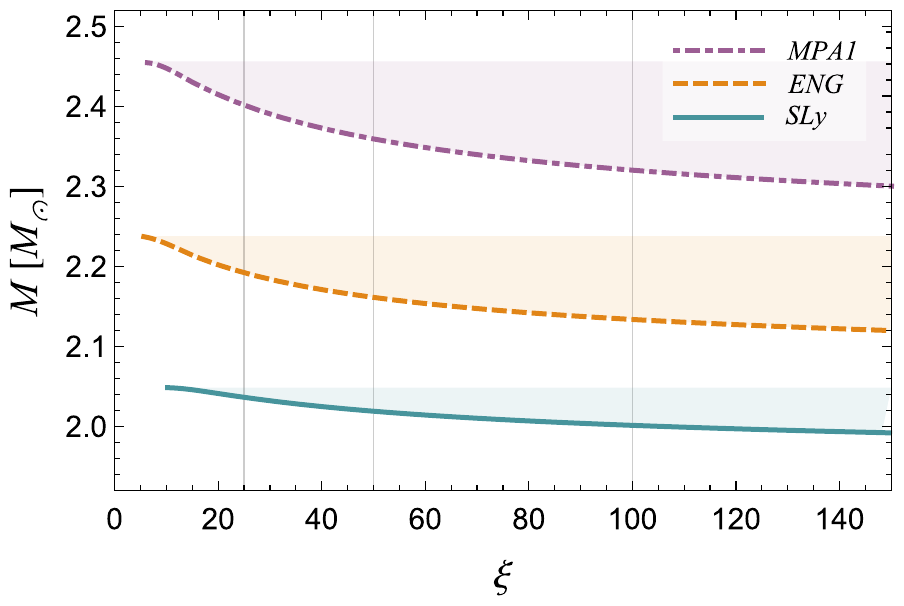}
\caption{Regions of the coupling parameter $\xi>0$ and NS masses where spontaneous scalarization can take place, for the three EoS considered in this work. Each region is cut at the maximum mass of a NS allowed, in GR, by that EoS. We highlight the values $\xi = 25$, $\xi = 50$ and $\xi = 100$ that will be considered subsequently.}
\label{fig:scalregion}
\end{figure}

Since the trace of the energy-momentum tensor, $T = 3p - \epsilon$, is typically negative (as energy density dominates over pressure), the scalarization effect would occur for $\xi < 0$ (so that $m_\textrm{eff}^2 <0$). Here, however, we will be mostly interested in studying pulsar-timing observables in the presence of positive values of the nonminimal coupling, since these are still unconstrained by astrophysical observations. In this case, in order to display the nontrivial phenomenology related to spontaneous scalarization, NSs must be sufficiently massive and compact that the trace of the energy-momentum tensor becomes positive in a region of their interior (so that $m_\textrm{eff}^2 <0$ in a sufficiently large region inside the star). In Ref.~\cite{Podkowka2018} the minimum compactness required for this property to hold was estimated to be $C \equiv M/R = 0.262^{+0.011}_{-0.017}$ (90\% confidence interval), with the error bars accounting for our ignorance on the nuclear EoS. All the EoS considered in this work allow for stars more compact than this threshold value. 

In Fig.~\ref{fig:scalregion} we show the regions of coupling parameters ($\xi>0$) and NS masses for which spontaneous scalarization takes place, for the three EoS considered in this work. To find these regions, we first solve the Tolman-Oppenheimer-Volkoff equations to construct a sequence of equilibrium configurations in GR and then determine whether unstable modes ($\delta \varphi \propto e^{\Omega t}$, $\Omega>0$) of Eq.~(\ref{eq:perturb}) can be found in that background. The lines delimiting the onset of instability coincide with the onset of spontaneous scalarization (as long as $\varphi_0 =0$) \cite{Pani2011,Mendes2015}. 
Notice that, although the realistic EoS considered in this work do not allow configurations that are compact enough to trigger this effect for the conformal coupling $\xi = 1/6$, more exotic structures could do so (see Fig.~3 of Ref.~\cite{Lima2013b} for the case of a thin shell).

\section{Pulsar-timing observables} \label{sec:observables}

Since the discovery of the first binary pulsar by Hulse and Taylor \cite{Hulse1975}, pulsar timing has become a major tool for testing GR (see \cite{Stairs2003,Wex2014} and references therein). The essence of pulsar timing lies in connecting the observed arrival time of the radio pulses to the proper time of emission. The resulting \textit{timing formula} is obtained from a succession of steps, as follows \cite{Blandford1976,Damour1986,Lorimer2005}. First, by relating proper time in the pulsar's rest frame to a coordinate system attached to the binary center of mass (CM), one ends up with contributions coming from the pulsar motion around the CM (the transverse Doppler shift), as well as from a varying gravitational redshift that depends on the relative distance between the pulsar and its companion. These effects are combined in the ``Einstein time delay'', usually expressed in terms of the measurable parameter $\gamma$. The next step consists in relating the coordinate time of emission to the coordinate time of arrival of a given pulse, by integrating along the null geodesics covered by the radiation. This picks up effects coming from dispersion in the interstellar medium, a geometrical contribution known as Roemer time delay, as well as from the Shapiro time delay due to the companion's gravitational well. Additional corrections are also accounted for, ranging from aberration effects due to the pulsar's rotation to corrections due to the motion of the Earth.  

At any instant, the orbit of each member of the binary system is tangent to a Keplerian ellipse (``osculating'' orbit), characterized by six orbital parameters---say, the semilatus rectum $p$, eccentricity $e$, longitude of pericenter $\omega$, time of pericenter passage $T_0$, inclination $i$, and angle of nodes $\Omega$. In order to account for deviations from Newtonian dynamics, these parameters are allowed to undergo secular variations: $x \to x + \dot{x} t$. In particular, the rate of periastron advance $\dot{\omega}$ and the rate of decay of the binary period $\dot{P}_b$ (derivable from the parameters above) are typically measurable.

In this work we will focus on the three classical pulsar-timing observables $\dot{\omega}$, $\gamma$, and $\dot{P}_b$, although many other post-Keplerian (PK) parameters can in principle be inferred from pulsar-timing observations \cite{Damour1992a}. A given theory of gravity will predict the value of these observables as a function of the Keplerian parameters and the masses of the binary components. If the masses ($m_p$ and $m_c$) are unknown, as often is the case, one can use the measurement of two of these PK parameters to infer $m_p$ and $m_c$, and perform a test of the gravitational theory with the third. This is typically portrayed by drawing, for each PK parameter, a level surface in the $m_p$-$m_c$ diagram corresponding to the measured value of that parameter. The theory is consistent with observations if the resulting curves all intersect at the same point, the binary component masses \cite{Wex2014}. 

In STTs, as in GR, the theoretical prediction for pulsar-timing observables is based on the PN description of the orbital motion. Explicitly, one has \cite{Damour1992,Damour1996}:
\begin{align}
\label{eq:omegadot}
\dot{\omega} &= \frac{3 n_b}{1-e^2} \frac{\mathsf{v}_b^2}{c^2} \left[
\frac{1-\alpha_p \alpha_c/3}{1+ \alpha_p \alpha_c} - \frac{m_p \beta_c \alpha_p^2 + m_c \beta_p \alpha_c^2}{6 M(1+\alpha_p \alpha_c)^2}\right], \\
\label{eq:gamma}
\gamma &= \frac{e}{n_b} \frac{m_c}{M(1+\alpha_c \alpha_p)} \frac{\mathsf{v}_b^2}{c^2} \left[ 1+ \alpha_c k_p + (1+\alpha_p \alpha_c) \frac{m_c}{M} \right],\\
\label{eq:Pbdot}
\dot{P}_b &= \dot{P}_\varphi^\textrm{monopole} + \dot{P}_\varphi^\textrm{dipole} + \dot{P}_\varphi^\textrm{quadrupole} + \dot{P}_{g*}^\textrm{quadrupole},
\end{align}
where $n_b\equiv 2\pi/P_b$, $M \equiv m_p + m_c$, and $\mathsf{v}_b \equiv (\mathcal{G} M n_b)^{1/3} $, with $\mathcal{G} \equiv G (1+\alpha_p\alpha_c)$. The quantities $\alpha_A$, $\beta_A$, and $k_A$, with the label $A\in\{p,c\}$ denoting the pulsar or its companion, are functions of the (Einstein-frame) stellar mass $m_A$, and will be defined in Sec.~\ref{sec:chargesdef} below. Equation (\ref{eq:Pbdot}) includes contributions related to monopole, dipole, and quadrupole scalar radiation, as well as the quadrupolar contribution from tensor waves familiar from GR. All terms are proportional to $(\mathsf{v}_b/c)^5$, and therefore of 2.5PN order, except from $\dot{P}_\varphi^\textrm{dipole}$, that contributes already at 1.5PN. Since this is the dominant contribution to the energy loss and will be important for our discussion, we write this term explicitly:
\begin{equation} \label{eq:Pbddip}
\dot{P}_\varphi^\textrm{dipole} = -\frac{2 \pi m_p m_c}{M^2(1+\alpha_p \alpha_c)} \frac{\mathsf{v}_b^3}{c^3} \frac{1 + e^2/2}{(1 - e^2)^{5/2}} (\alpha_p - \alpha_c)^2.
\end{equation}
For comparison,
\begin{equation}\label{eq:Pbdquad}
\dot{P}_{g*}^\textrm{quad} = -\frac{192 \pi m_p m_c}{5 M^2 (1+\alpha_p \alpha_c)} \frac{(\mathsf{v}_b/c)^5}{(1-e^2)^{7/2}} \left( 1+ \frac{73}{24}e^2 + \frac{37}{96}e^4\right).
\end{equation}
The expressions for the other terms can be found, e.g., in Refs.~\cite{Damour1992,Mirshekari2013}. 

\subsection{Scalar charges}\label{sec:chargesdef}

Crucially, in the pulsar timing observables (\ref{eq:omegadot})-(\ref{eq:Pbdot}) we find the appearance of the ``gravitational form factors'' (in the terminology of Ref.~\cite{Damour1996}) or ``scalar charges'' (in the terminology of Ref.~\cite{Anderson2019}) $\alpha_A = \alpha_A(m_A)$, $\beta_A = \beta_A(m_A)$, and $k_A = k_A(m_A)$ (with $A \in \{p,c\}$). In this section we briefly review their definition, the rationale behind their appearance in pulsar-timing observables, and how to compute them in practice.

It is well known that, in GR, a body's orbital dynamics is governed by integral quantities such as its mass and spin, with corrections due to shape and internal structure appearing only at high PN orders.
For this reason, complex bodies can often be well approximated by point masses (insofar as their orbital motion is concerned). In STTs, the value of the scalar field around the star determines the strength of the effective gravitational coupling, and therefore influences its global properties. This feature can be effectively incorporated into a PN description based on point masses by allowing the mass of each body to be field dependent: For an $N$-body system, the action is taken to be \cite{Eardley1975}
\begin{equation}
S_m = - \sum_{A=1}^N \int m_A (\varphi_A) d\tau^*_A.
\end{equation}
It is worthwhile to emphasize the dual role played by the function $m_A(\varphi_A)$ above. If one assumes that the inter-body distance $D$ between the binary components is much larger than their typical size $R$, and allow the ratio $R/D$ to shrink to zero, then $m_A(\varphi_A)$ expresses the stellar mass in terms of the scalar field value at the star's location. This is the ``outer'', PN perspective. However, the  function $m_A(\varphi_A)$ is not determined self-consistently within the PN scheme, but is assumed to be known from the matching to the ``inner'' problem, where the stellar structure is determined. From this ``inner'' perspective, the point particle limit corresponds to the matching sphere becoming infinitely large, and from this point of view the function $m_A(\varphi_A)$ denotes the mass the star has when the asymptotic value of the scalar field is $\varphi_A$.

Once the field is expanded around its (cosmological) asymptotic value in the PN approximation, the effect of a field-dependent mass is encoded in the asymptotic value of its derivatives. In particular, one defines:
\begin{equation}\label{eq:alpha}
\alpha_A \equiv \left. \frac{d \log m_A}{d\varphi_A} \right|_{\varphi_0}, 
\end{equation}
\begin{equation}\label{eq:beta}
\beta_A \equiv \left. \frac{d\alpha_A}{d\varphi_A}  \right|_{\varphi_0},
\end{equation}
and so on. 
The scalar charges above (closely related to the ``sensitivities'' used in the Jordan-frame description \cite{Mirshekari2013}) are the ones appearing at the Newtonian and post-Newtonian levels, relevant for the derivation of the pulsar timing observables (\ref{eq:omegadot})-(\ref{eq:Pbdot}).

The additional scalar charge $k_A$ entering the Einstein time delay (\ref{eq:gamma}) is defined as 
\begin{equation}\label{eq:kA}
k_A \equiv - \left.  \frac{\partial \log I_A}{\partial \varphi_A}  \right|_{\varphi_0},
\end{equation}
where $I_A$ is the moment of inertia of star $A$. It appears in the computation of the parameter $\gamma$ when relating the intrinsic time of the pulsar clock to the proper time in a local inertial frame around the pulsar, due to the fact that, in STTs, the moment of inertia of the star depends on the local scalar field environment---which may fluctuate in a binary system, causing the angular velocity to fluctuate as well \cite{Damour1996}. 

In order to explain how the scalar charges $\alpha_A$, $\beta_A$, and $k_A$ are computed, we now revert attention to the inner problem of an isolated, slowly rotating star, which is a suitable approximation for most of the observed pulsars. We therefore consider the spacetime of a slowly rotating body, with metric \cite{Hartle1967}
\begin{align}\label{eq:slowrot}
ds_*^2 &= - e^{\nu (r)} dt^2 + e^{\lambda(r)} dr^2 + r^2 d\theta^2 + r^2 \sin^2 \theta d\phi^2 \nonumber \\
& - 2 \varpi(r) r^2 \sin^2\theta dt d\phi,
\end{align}
where $\varpi(r)$ accounts for frame-dragging effects and is the only $O(\Omega)$ correction to the static case, with $\Omega$ denoting the star's angular velocity (see Refs.~\cite{Doneva2013,Pani2014} for generalizations). 
The differential equations governing the metric functions $\nu(r)$, $\lambda(r)$, $\varpi(r)$, scalar field $\varphi(r)$, and fluid variable $p(r)$ can be found, e.g., in Refs.~\cite{Damour1996,Sotani2012}, and are reproduced in Appendix \ref{sec:numerics} for the sake of completeness. By imposing regularity conditions at $r=0$, asymptotic flatness at spatial infinity, as well as the condition $\varphi(r) \overset{r \to \infty}{\longrightarrow} \varphi_A$, one can integrate these equations to obtain a one-parameter family of solutions (labeled, say, by the central value of the pressure). 
The scalar charges in Eqs.~(\ref{eq:alpha})-(\ref{eq:kA}) measure the change of the stellar properties as $\varphi_A$ changes, while keeping fixed the baryon mass inside the star ($\bar{m}_A = \int \rho \sqrt{-g} u^0 d^3x $, where $\rho$ is the rest-mass density).

Global properties of the NS can be extracted from the asymptotic behavior of the metric functions. The total mass $m_A$ is obtained from the $1/r$ behavior of $e^{\nu(r)} \overset{r\to\infty}{\longrightarrow} 1 - 2 m_A/r + O(1/r^2)$. The total angular momentum $J_A$ is obtained from the $1/r^3$ behavior of $\varpi(r) \overset{r\to\infty}{\longrightarrow} 2J_A/r^3 + O(1/r^4)$, and enables the computation of the star's moment of inertia: $I_A = J_A/\Omega$. Finally, from the leading $1/r$ contribution to the scalar field at spatial infinity, $\varphi \overset{r\to\infty}{\longrightarrow} \varphi_A + \omega_A/r + O(1/r^2)$, one can directly obtain the scalar charge $\alpha_A$ through $\alpha_A = -\omega_A / m_A$. That the quantity computed in this way is equivalent to the definition in Eq.~(\ref{eq:alpha}) is shown in the Appendix A of Ref.~\cite{Damour1992}. 

\begin{figure*}[t]
\includegraphics[width=0.8\textwidth]{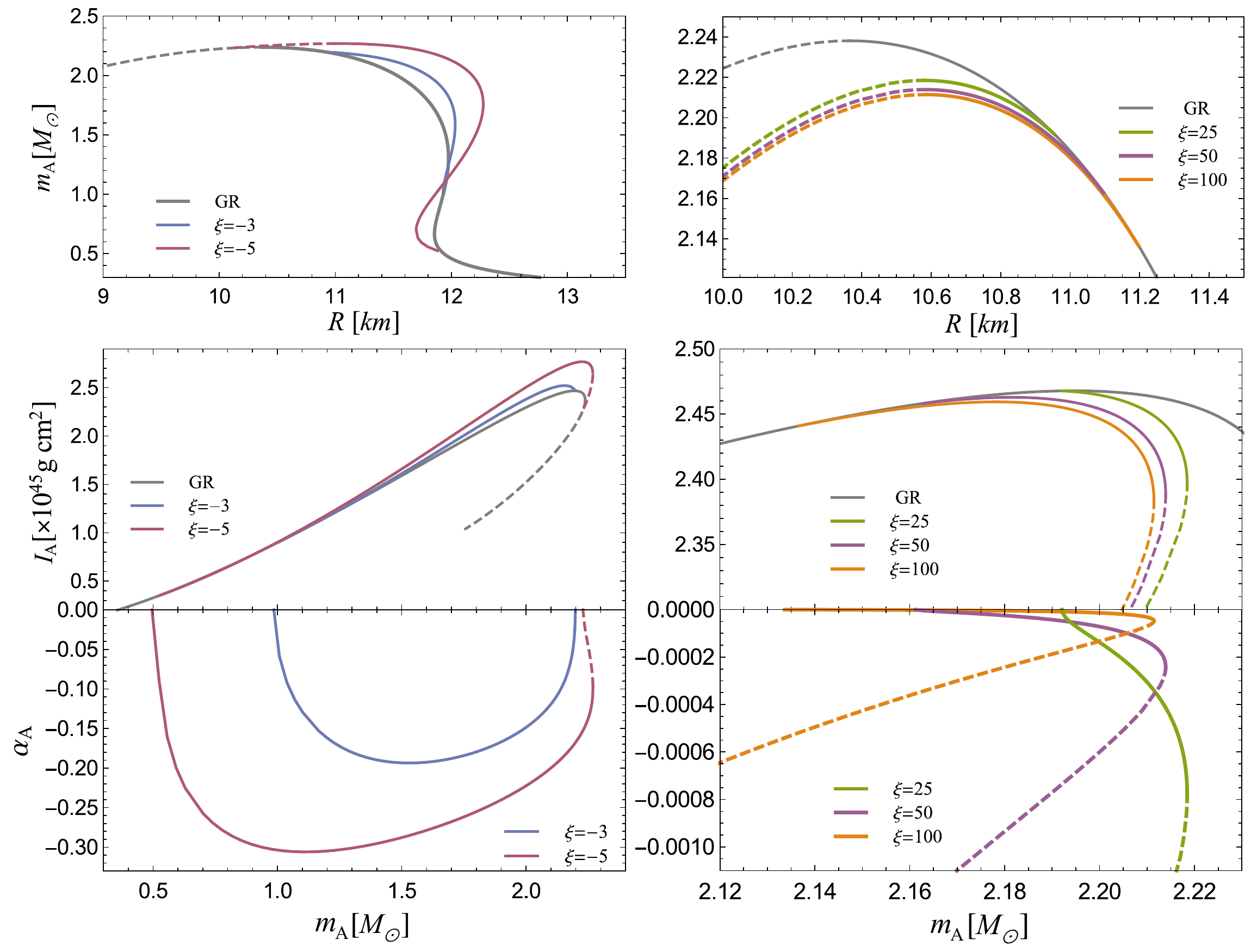}
\caption{Total mass $m_A$, moment of inertia $I_A$ and scalar charge $\alpha_A$ of a sequence of equilibrium solutions describing NSs with the ENG EoS in the presence of a scalar field with SNMC $\xi = -3$ and $-5$ (left column) and $\xi = 25, 50$, and $100$ (right column). The asymptotic value of the scalar field was fixed to zero. The dashed part of each curve denotes their hydrodynamically unstable piece.}
\label{fig:MalphaI}
\end{figure*}

Contrary to the computation of $\alpha_A$, which can be done straightforwardly, computing $\beta_A$ and $k_A$ is more involved, and here we proceed as follows. 
First, we construct three sequences of equilibrium solutions, for $\varphi_A^{(0)} = \varphi_0$, $\varphi_A^{(+)} = \varphi_0 + \Delta \varphi$ and $\varphi_A^{(-)} = \varphi_0 - \Delta \varphi$, storing, for each value of the central pressure, the total mass $m_A$, baryonic mass $\bar{m}_A$, and moment of inertia $I_A$ of the resulting solution. This data is then used to define, by interpolation, functions $m^{(i)}_A(\bar{m}_A)$, $\alpha^{(i)}_A(\bar{m}_A)$, and $I^{(i)}_A(\bar{m}_A)$, with $i \in \{0,+,-\}$, from which we can compute the scalar charges by a finite difference approximation to the derivative operators in Eqs.~(\ref{eq:beta}) and (\ref{eq:kA}). For our purposes, we find it enough to use a central stencil for the derivative operator, setting, for instance,
\begin{equation}
k_A(\bar{m}_A) \approx \frac{1}{I^{(0)}_A(\bar{m}_A)}\frac{I^{(+)}(\bar{m}_A) - I^{(-)}(\bar{m}_A)}{2\Delta \varphi}.
\end{equation}
More details on the numerical procedure and the accompanying errors can be found in Appendix~\ref{sec:numerics}. Note also that a bank of scalar charges, as well as a thorough account of the procedure to compute them, was recently provided in  Ref.~\cite{Anderson2019}. The authors did not, however, explore the possibility of spontaneous scalarization for $\xi>0$, which is our main focus here.

\begin{figure*}[t]
\includegraphics[width=0.9\textwidth]{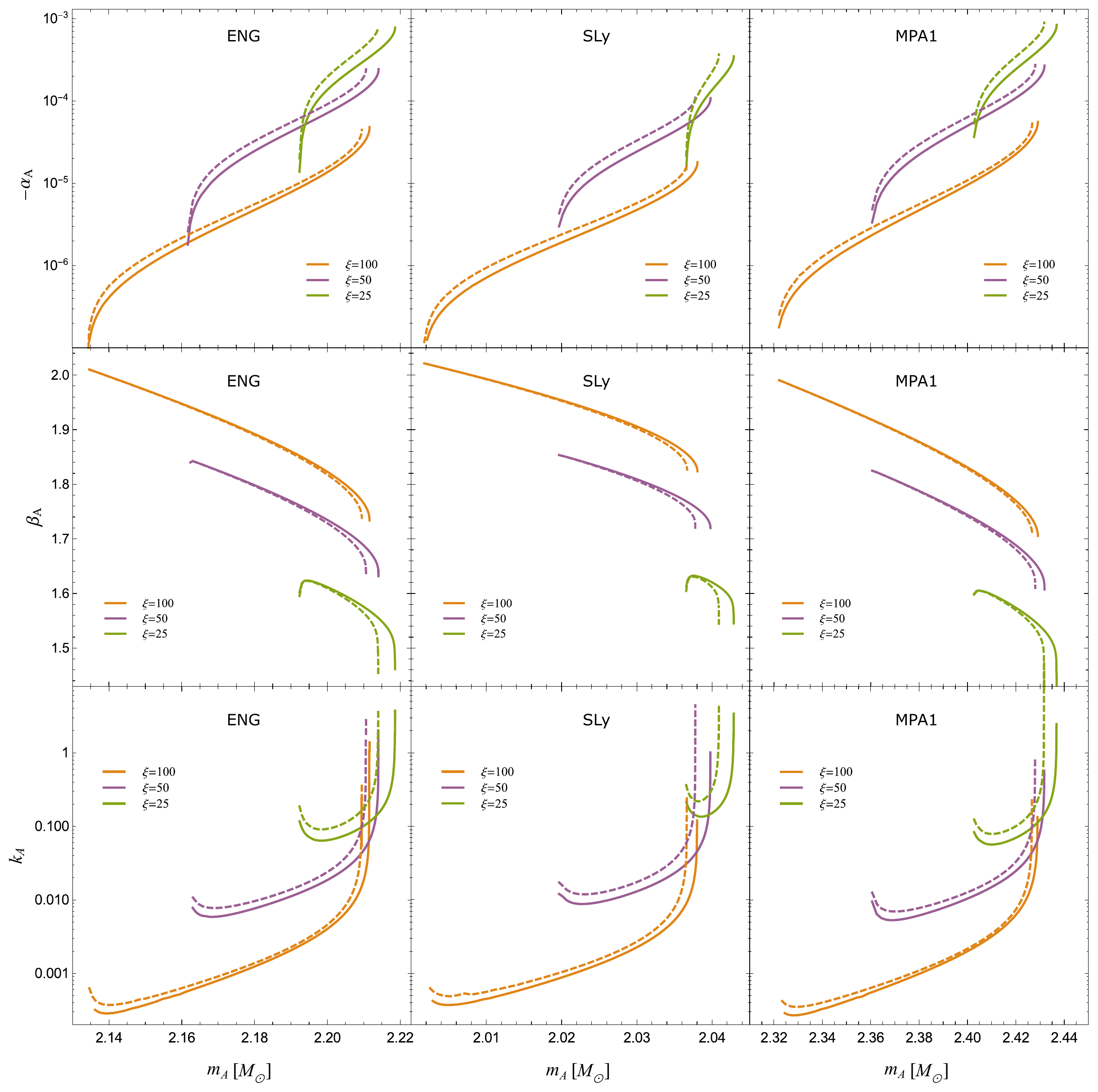}
\caption{Scalar charges $\alpha_A$, $\beta_A$, and $k_A$ as a function of the stellar mass for $\xi = 25$, 50, and 100, and three nuclear EoS: ENG, SLy, and MPA1. Solid lines refer to the SNMC, while dashed lines of the same color correspond to the HC model, with the same value of $\xi$.}
\label{fig:abk}
\end{figure*}

\section{Results} \label{sec:results}

\subsection{Scalar charges} \label{sec:charges}

In order to establish a point of comparison between the cases where $\xi$ is negative and positive, in Fig.~\ref{fig:MalphaI} we show the behavior of the total mass $m_A$, moment of inertia $I_A$, and scalar charge $\alpha_A$ for a sequence of equilibrium solutions with $\xi = -3$ and $-5$ and $\xi = 25, 50,$ and $100$. The asymptotic value of the scalar field is taken to be $\varphi_0 = 0$.
As anticipated in Sec.~\ref{sec:scalarization} (see Fig.~\ref{fig:scalregion}), scalarized solutions for $\xi >0$ only exist for the most massive and compact stars, while scalarization happens for $\xi <0$ in a much wider range of masses. From the first row of Fig.~\ref{fig:MalphaI}, we see that as $|\xi|$ increases the maximum mass of scalarized solutions increases when $\xi <0$ and decreases when $\xi >0$. This is consistent with the interpretation of $G_\textrm{eff} = G/F(\Phi)$ as an effective gravitational coupling. As $\xi>0$ increases, so does $G_\textrm{eff}$: gravity becomes stronger in the presence of the scalar field and less massive scalarized stars can be supported without undergoing gravitational collapse. The opposite happens for $\xi <0$.
For a given mass, the second row of Fig.~\ref{fig:MalphaI} shows that the moment of inertia is larger (smaller) than in GR when $\xi < 0 $ ($>0$). This can also be understood as the effect of gravity becoming weaker (stronger) in each of these cases, and the stellar size becoming typically larger (smaller) than in GR when a nontrivial scalar field profile is present.

Since the field equations (\ref{eq:metric_eq}) and (\ref{eq:phi_eq}) are invariant under the transformation $\varphi \to - \varphi$ for the coupling functions we consider, one always finds two twin scalarized solutions related by the aforementioned transformation, and with opposite scalar charges $\alpha_A$. This is true as long as $\varphi_0 = 0$, as in Fig.~\ref{fig:MalphaI}, otherwise the boundary condition breaks the reflection symmetry of the solution.
In the third row of Fig.~\ref{fig:MalphaI} we show the scalar charge $\alpha_A$ for solutions with a positive scalar field profile. 

Two notable differences are found in the $\xi>0$ case (right panel), compared to when $\xi<0$ (left panel), namely that the magnitude of $\alpha_A$ (i) is now much smaller and (ii) typically decreases as $|\xi|$ increases.
Both have to do with the fact that the scalar field tends to be amplified in the stellar region where $\xi T > 0$ and suppressed in the region where $\xi T <0$. For any realistic EoS, the trace of the energy momentum can be positive only in a small region in the stellar interior; therefore, a scalar field with $\xi >0$ is amplified in this inner region, but is necessarily suppressed in the outer layers of the star. As a consequence, although the central value of the scalar field may increase with increasing $\xi$, the scalar charge, measured asymptotically, ends up being smaller. In Appendix \ref{sec:toymodel} we present a simple, analytically solvable, toy model that illustrates these points. 

Figure \ref{fig:abk} shows the scalar charges $\alpha_A$, $\beta_A$, and $k_A$ for $\xi = 25$, 50, and 100, for both the SNMC and the HC models, and for three nuclear EoS describing the NS fluid. From the first row we see that the properties of $\alpha_A$ described in the preceding paragraph are not altered by the nuclear EoS. Indeed, the magnitude of all scalar charges is only mildly influenced by the EoS, the main effect of which is to change the range of masses where scalarization takes place. 

\begin{figure}[t]
\includegraphics[width=0.9 \linewidth]{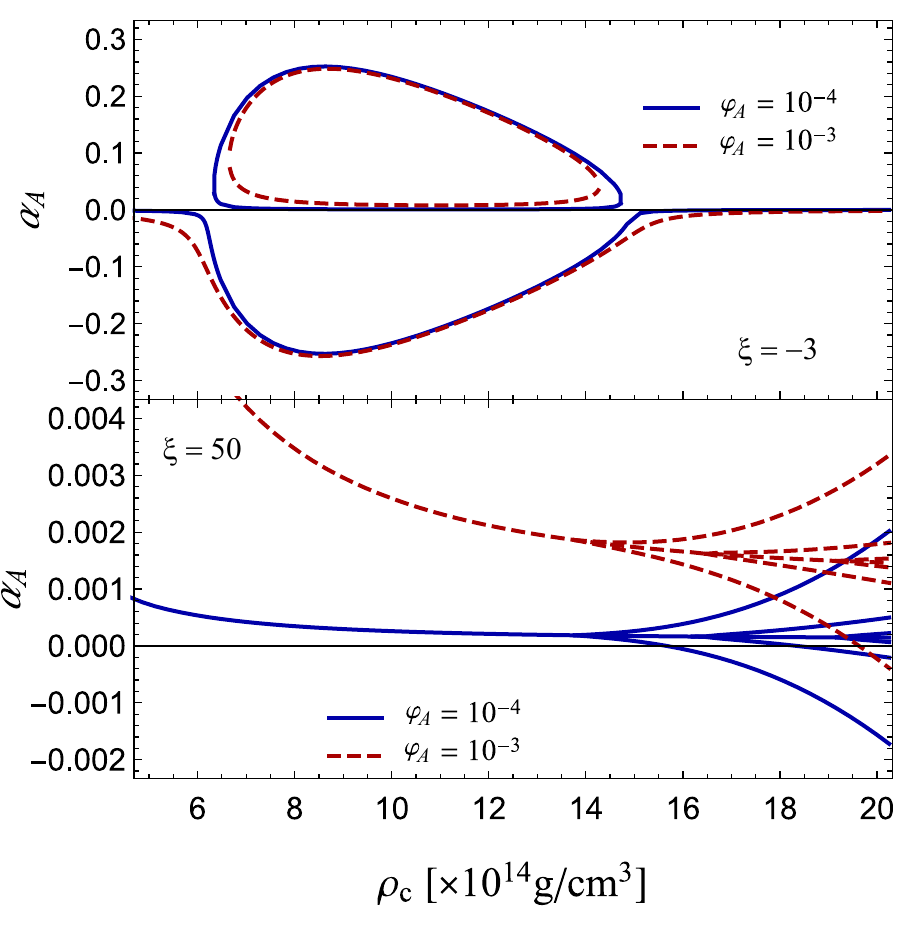}
\caption{Scalar charge $\alpha_A$ as a function of the central rest-mass density $\rho_c$ for the HC model with $\xi = -3$ (upper panel) and $\xi = 50$ (lower panel), and two asymptotic values of the scalar field: $\varphi_A = 10^{-4}$ and $\varphi_A = 10^{-3}$. The ENG EoS was assumed. All families of equilibrium solutions are shown, and a change in the number of solutions is noted at several critical densities (note, however, that not all the solutions in the $\xi>0$ case are stable, as shown in the Supplemental Material of Ref.~\cite{Mendes2018}).}
\label{fig:alphaphi0}
\end{figure}

The second row of Fig.~\ref{fig:abk} shows the scalar charge $\beta_A = (d\alpha_A/d\varphi_A)_{\varphi_0}$ as a function of the stellar mass. This is a positive, order-of-unity quantity, which increases with increasing $\xi$. The fact that $\beta_A$ is positive (and $\alpha_A$ negative) means that $\alpha_A$ receives a positive increment when the asymptotic value of the scalar field $\varphi_A$ increases, and therefore typically diminishes in absolute value (as long as $\varphi_A$ is close to 0). This is opposite to what happens when $\xi <0$, in which case $\beta_A <0$ and $\alpha_A$ increases in absolute value with increasing $\varphi_A$. This is shown more explicitly in Fig.~\ref{fig:alphaphi0}, where the scalar charge $\alpha_A$ is represented for $\varphi_A = 10^{-4}$ and $10^{-3}$, for $\xi = 50$ and $\xi = -3$. 

Interestingly, the features described above can have implications for scalarization in dynamical situations.
``Dynamical scalarization'' was first observed in binary NS simulations in STTs \cite{Barausse2013}, wherein NSs that were not compact enough to scalarize in isolation suddenly developed a large scalar charge once their separation became sufficiently small, making the coalescence proceed in a faster timescale. It was then understood to be due to a kind of feedback mechanism \cite{Palenzuela2014,Sennett2016}. 
Indeed, as discussed before, from the PN perspective the mass and other stellar properties are functions of the local scalar field value at the star's worldline, which is influenced by the presence of the companion star. In STTs with $\xi <0$, if the ambient scalar field value grows, it induces a growth in $\omega_A = -m_A \alpha_A$ (see Fig.~\ref{fig:alphaphi0}). This, in turn, increases the local value of the scalar field at the companion's location (given by $\varphi_B \sim \varphi_0 + \omega_A/r$, where $r$ is the separation distance), and the positive feedback proceeds until a fixed-point is reached. Although exploring this in detail is beyond the scope of the present work, we can anticipate that the opposite behavior is likely to occur in STTs with $\xi >0$: In this case $\beta_A >0$ and $\omega_A$ decreases with increasing $\varphi_A$; therefore the feedback mechanism present in the $\xi <0$ case will most likely be absent or reversed.

Finally, let us come back to the last row of Fig.~\ref{fig:abk}, where $k_A$ [cf.~Eq.~(\ref{eq:kA})] is shown as a function of the stellar mass. The typical values of $k_A$ are seen to decrease with increasing $\xi$, with a large spike close to the maximum allowed mass. This is a somewhat similar behavior to the $\xi < 0$ case (see, e.g., Fig.~4 in Ref.~\cite{Anderson2019}), although the typical values in Fig.~\ref{fig:abk} are quite small due to the relatively large values of $\xi$ we consider. In all cases, the HC reproduces well the qualitative features of the SNMC.

\subsection{Implications for pulsar-timing observables} \label{sec:pulsardata}

Having understood the properties of the scalar charges $\alpha_A(m_A)$, $\beta_A(m_A)$, and $k_A(m_A)$, we now turn to the pulsar timing observables $\dot{\omega}$, $\gamma$, and $\dot{P}_b$, given in STTs as in Eqs.~(\ref{eq:omegadot})-(\ref{eq:Pbdot}). 

Pulsar-timing observations can be used to put stringent bounds on $\xi < 0$ \cite{Damour1996,Freire2012,Shao2017,Anderson2019,Anderson2019a}. 
A crucial feature that makes this possible is that the scalar charge $\alpha_A$, which enters in all pulsar-timing observables, increases in magnitude as $\xi$ becomes more negative, making deviations from GR stronger. As a consequence, one is able to exclude the entire range of negative couplings up to a certain value. 

The situation for $\xi>0$ is much more permissive. On the one hand, since the scalar charge $\alpha_A$ typically decreases in magnitude as $\xi>0$ increases (cf.~Fig.~\ref{fig:abk}; see also Fig.~8 of Ref.~\cite{Anderson2017}), deviations of pulsar-timing observables from GR are suppressed for large values of the nonminimal coupling. On the other hand, as we decrease the value of $\xi$ in search of larger scalar charges, the range of NS masses allowing for spontaneous scalarization gets progressively smaller (cf.~Fig.~\ref{fig:scalregion}). Thus, STTs with $\xi>0$ tend to pass pulsar-timing tests, with the possible exception of a small range of couplings and NS masses. 

Having described the global picture, let us give a few more details about each of the observables $\dot{\omega}$, $\gamma$, and $\dot{P}_b$. 
Since $\beta_A = O(1)$ for $\xi > 0$ (cf.~Fig.~\ref{fig:abk}), we can see from Eq.~(\ref{eq:omegadot}) that corrections to $\dot{\omega}$ are $O(\alpha_A^2)$, and therefore negligible due to the smallness of the scalar charge $\alpha_A$ in this case.
The same is true for the Einstein time delay $\gamma$, except for the term proportional to $\alpha_c k_p$ in Eq.~(\ref{eq:gamma}). Since $k_A$ seems to diverge in limit of the maximum NS mass (cf.~Fig.~\ref{fig:abk}), this term still could give a non-negligible contribution as long as the companion's charge $\alpha_c$ is not vanishingly small. This would be restricted, however, to an exceptionally thin range of NS masses. As for the rate of decay of the orbital period, $\dot{P}_b$, it is enough to compare the contribution in Eq.~(\ref{eq:Pbddip}) coming from dipolar scalar radiation to the usual quadrupolar contribution due to tensor waves, Eq.~(\ref{eq:Pbdquad}). Although the first one dominates the second by a factor of $(c/\mathsf{v}_b)^2$, it is also suppressed by a factor of $(\alpha_p - \alpha_c)^2$. Typical orbital velocities in binary systems observed through pulsar-timing techniques are of the order of $(\mathsf{v}_b/c) \sim 10^{-3}$. Interestingly, this is of the same order of magnitude as the largest values of $|\alpha_A|$ shown in Fig.~\ref{fig:MalphaI}. Therefore, it might still be possible to probe STTs with $\xi \lesssim 25$ with the radiation emitted by the most massive pulsars. However, as emphasized above, the range of masses allowing for spontaneous scalarization gets narrower as $\xi$ decreases, and such effects, if present, would be restricted to very special systems. For the largest ranges of NS masses and nonminimal couplings, STTs with $\xi>0$ would still evade the sharp knife of pulsar-timing tests.

\section{Conclusion} \label{sec:conclusion}

It is well known that general relativity has passed with flying colors all strong-field tests imposed by pulsar-timing observations. Moreover, these observations have stripped many modified theories of gravity of a large portion of their parameter space, confining their predictions to the close vicinity of GR \cite{Will1993}. This is the case, in particular, of a class of scalar-tensor theories of gravity including the case of a massless scalar field with the standard, $\xi R \Phi^2$, nonminimal coupling to gravity. In these theories a tachyonic-like instability develops for sufficiently compact stars and $\xi\lesssim - 2.2$, leading to the development of a scalar cloud around the star (see Sec.~\ref{sec:framework}). If the scalar field is thus activated around a member of a binary system, it drives the emission of dipole scalar radiation, contributing to a steeper decrease of the orbital period in time. Since this effect---as well as the changes introduced in other observable quantities (see Sec.~\ref{sec:observables})---is incompatible with pulsar-timing data, almost the entire range of couplings allowing for spontaneous scalarization has now been excluded \cite{Freire2012,Anderson2019a}.

Recently, it has been advocated that a similar scalarization effect might occur for positive values of the nonminimal coupling $\xi$, around the most massive neutron stars found in Nature, as long as they are also sufficiently compact \cite{Mendes2016,Podkowka2018} (which depends on the still unknown nuclear equation of state). This seems to open the possibility of using measurements of the most massive observed NS, such as the pulsar PSR J0348+0432 \cite{Antoniadis2013} or possibly PSR B1957+20 \cite{vanKerkwijk2011}, in order to also probe this range of couplings. 

In this work we carried out a study of the scalar charges $\alpha_A$, $\beta_A$, and $k_A$, which determine the behavior of the main post-Keplerian pulsar timing observables, $\dot{\omega}$, $\gamma$, and $\dot{P}_b$, in STTs with $\xi >0$ (see Sec.~\ref{sec:results}). We find that the scalar charges differ remarkably in the $\xi >0$ and $\xi<0$ cases. In particular, the main scalar charge $\alpha_A$ governing Newtonian and post-Newtonian deviations from GR is suppressed as $\xi>0$ becomes large, while the range of masses allowing for spontaneous scalarization decreases as $\xi>0$ becomes smaller. As discussed in detail in Sec.~\ref{sec:results}, this indicates that STTs with $\xi >0$, even in the presence of spontaneous scalarization, are able to pass pulsar-timing tests, with the possible exception of an exceedingly narrow range of couplings and NS masses. Moreover, the contrasting properties exhibited by the scalar charge $\beta_A$ in the $\xi>0$ case suggest that the feedback mechanism responsible for the effect of dynamical scalarization might not be present for positive values of the nonminimal coupling. It remains an interesting research avenue to determine which NS properties are most sensitive to a scalar field background in this regime of small scalar charges (e.g.~the NS oscillation spectrum as recently suggested by Ref.~\cite{Mendes2018}).

\acknowledgments
We are glad to thank Gabriel Vidal for his involvement in the early phases of this project, and N\'estor Ortiz for valuable comments on the manuscript.
R.M. acknowledges partial financial support by the National Council for Scientific and Technological Development--CNPq. T.O. acknowledges financial support from funding agencies CAPES and FAPERJ. 

\appendix

\section{A simple toy model} \label{sec:toymodel}

In order to gain some intuition about the behavior of the scalar charge $\alpha_A$, we consider a simpler version of Eq.~(\ref{eq:phi_eq}) \cite{Damour1993}, where we neglect the metric curvature and the details of the coupling function, keeping only its linear piece:
\begin{equation}
\Delta \varphi = - \epsilon \kappa^2 \varphi.
\end{equation}
Here $\Delta \equiv d^2/dr^2 + (2/r) d/dr$, $\kappa \equiv \sqrt{8 \pi |\xi T|}$, and $\epsilon = \textrm{sign} (\xi T)$. Moreover we will assume that $\kappa$ is constant throughout the star, but that $\epsilon$ can change sign, according to
\begin{equation} \label{eq:epsilon}
\epsilon = 
\begin{cases}
+ 1, & \mbox{if } \; 0 \leq r \leq r^+ \\
- 1, & \mbox{if } \; r^+ < r \leq R
\end{cases}
\end{equation}
where $R$ is the stellar radius and $0 \leq r^+ \leq R$. The case most often considered in the literature, i.e., $\xi <0$ and $T<0$ throughout the star, would correspond to $r^+ = R$. On the other hand, we are most interested here in the case where $\xi >0 $ and the trace $T$ changes sign inside the star, as captured by the general form (\ref{eq:epsilon}).
Taking $\kappa$ to be constant, as we do for simplicity, implies the additional assumption that the typical strength (say, the mean value) of $T>0$ in the region $0 \leq r \leq r^+$ is comparable with the typical value of $T<0$ in the region $r^+ < r \leq R$.

The general solution to this problem is
\begin{equation}
\varphi =
\begin{cases}
\varphi_c \frac{\sin(\kappa r)}{\kappa r}, & \mbox{if } \; 0 \leq r \leq r^+ , \\
A \frac{\sinh(\kappa r)}{\kappa r} + B \frac{\cosh (\kappa r)}{\kappa r}, & \mbox{if } \; r^+ < r \leq R , \\
\varphi_0 + \omega/r, & \mbox{if } \; r > R ,
\end{cases}
\end{equation}
where $\varphi_c$, $A$, $B$, and $\omega$ can be explicitly computed in terms of $\varphi_0$, $\kappa$, $r^+$, and $R$ by imposing continuity of the field and its derivative across $r=r^+$ and $r=R$. In particular, 
\begin{equation}\label{eq:phicfull}
\varphi_c = \frac{\varphi_0}{\sin(\kappa r^+) \sinh(\kappa r^-) + \cos(\kappa r^+) \cosh(\kappa r^-)}
\end{equation}
and
\begin{equation}\label{eq:omegafull}
\omega = - \varphi_0 R + \frac{\varphi_0}{\kappa} \frac{\tan(\kappa r^+) + \tanh(\kappa r^-)}{1 + \tan(\kappa r^+) \tanh(\kappa r^-)},
\end{equation}
where we defined $r^- \equiv R - r^+$. It is worth writing explicitly two particular cases: If $r^+ = 0$, we have
\begin{equation}\label{eq:case1}
\varphi_c = \frac{\varphi_0}{\cosh(\kappa R)}, \quad \omega = - \varphi_0 R \left( 1 - \frac{\tanh(\kappa R)}{\kappa R} \right),
\end{equation}
while if $r^- = 0$ we get
\begin{equation}\label{eq:case2}
\varphi_c = \frac{\varphi_0}{\cos(\kappa R)}, \quad  \omega = - \varphi_0 R \left( 1 - \frac{\tan(\kappa R)}{\kappa R} \right).
\end{equation}
In the first case, Eq.~(\ref{eq:case1}), the field is suppressed with respect to its asymptotic value and $|\omega|$ is bounded by $|\varphi_0| R$, going to zero as $\varphi_0 \to 0$. This would be the picture if, say, $\xi >0$ and $T<0$ throughout the star. In the second case, Eq.~(\ref{eq:case2}), the field is amplified with respect to its asymptotic value and $|\omega|$ is enhanced with respect to $|\varphi_0| R$. Indeed, these quantities may have a nonzero limit even when $\varphi_0 \to 0$, as long as $\kappa R = \pi/2$. 
This gives a heuristic picture of the spontaneous scalarization effect when $\xi < 0 $ \cite{Damour1993}.

Take now the full expressions (\ref{eq:phicfull}) and (\ref{eq:omegafull}), which are the relevant ones when $\xi >0$. We see that there is a competition between enhancement and suppression effects. In particular, the onset of scalarization, which was determined by the condition $\cot(\kappa R) = 0$ for Eq.~(\ref{eq:case2}), is delayed to higher values of $\kappa R$, with the relevant condition becoming $\cot(\kappa r^+) = - \tanh (\kappa r^-)$. 
Moreover, a number of features displayed by the scalar charge $\alpha_A$ for scalarized solutions (cf.~Sec.~\ref{sec:charges}) are already exhibited in the expressions above (outside the scalarization regime). For instance, the $1/\kappa$ dependency in Eq.~(\ref{eq:omegafull}) indicates that $|\omega|$ decreases with increasing $\kappa$, except near the resonances mentioned above. This is in contrast with the case (\ref{eq:case2}), where $|\omega|$ increases with $\kappa$.

\section{Numerical setup} \label{sec:numerics}

With the metric ansatz (\ref{eq:slowrot}), the field equations (\ref{eq:metric_eq}) and (\ref{eq:phi_eq}) yield
\begin{align}
m' &= 4 \pi r^2 F^{-2} \epsilon + \frac{1}{2} r(r-2m) (\varphi')^2, \label{eq:meq} \\
\nu' & = \frac{8\pi r^2 F^{-2} p}{r-2m} + r(\varphi')^2 + \frac{2m}{r(r-2m)}, \\
\varphi'' &= \frac{4\pi r F^{-2}}{r-2m} \left[ \alpha (\epsilon - 3 p) + r \varphi' (\epsilon - p) \right] - \frac{2(r-m)}{r(r-2m)} \varphi',  \\
p' &= - (\epsilon + p) \left[ \frac{4\pi r^2 F^{-2} p}{r-2m} + \frac{r(\varphi')^2}{2} + \frac{m}{r(r-2m)} + \alpha \varphi' \right],  \\
\varpi'' &= \varpi' \left[ -\frac{4}{r} + r (\varphi')^2 + \frac{4\pi r^2 (\epsilon + p)}{F^2 (r-2m)}  \right] \nonumber \\
 & + \frac{16\pi r (\epsilon + p)}{F^2 (r-2m)} (\Omega - \varpi), \label{eq:omegaeq}
\end{align}
where primes denote radial derivatives, $\Omega \equiv u^\phi/u^t$ is the fluid angular velocity, and we have set $e^{\lambda(r)} \equiv 1/(1-2m(r)/r)$. When supplemented by a choice of EoS, the set of differential equations above can be integrated numerically by standard methods. The relevant boundary conditions to the problem are regularity at $r=0$, which requires that $m(0) = 0$, $\varpi'(0) = 0$, and $\varphi'(0) = 0$, and asymptotic flatness, which requires that $\lim_{r\to \infty} \nu(r) = 0$, $\lim_{r\to \infty} \varpi(r) = 0$ and $\lim_{r\to \infty} \varphi(r) = \varphi_A$. The condition $p(R) = 0$ determines the stellar radius $R$. As described in Sec~\ref{sec:chargesdef} the stellar mass $m_A$ and angular momentum $J_A$ are determined by the asymptotic behavior of the metric functions. Alternatively, Eqs.~(\ref{eq:meq})-(\ref{eq:omegaeq}) can be integrated up to the stellar radius and matched to the known form of the exterior solution \cite{Damour1996}, from which global quantities like $m_A$ and $J_A$ can be inferred.

As described in Sec.~\ref{sec:charges}, the computation of the scalar charges, 
\begin{equation} \label{eq:charges}
\alpha_A = \frac{d \log m_A}{d\varphi_A}, \quad
\beta_A = \frac{d \alpha_A}{d\varphi_0}, \quad
k_A = - \frac{d \log I_A}{d\varphi_0}, \quad
\end{equation}
is somewhat involved, since the derivatives must be evaluated for a fixed value of the baryonic mass. For this purpose, given an EoS and a value for $\xi$, we construct three sequences of equilibrium solutions, with the asymptotic value of the scalar field given by $\varphi_A^0 = \varphi_0$, $\varphi_A^+ =\varphi_0 + \Delta \varphi$, and $\varphi_A^- =\varphi_0 - \Delta \varphi$. For most of the results presented in this work we employ $\varphi_0 = 0$ and $\Delta \varphi = 0.0005$, except for Fig.~\ref{fig:alphaphi0} where different values of $\varphi_0$ are used. From the data thus generated, we construct by interpolation the functions $m_A^{(i)}(\bar{m}_A)$, $\alpha_A^{(i)} (\bar{m}_A)$, and $I_A^{(i)}(\bar{m}_A)$, with $i \in \{0,+,-\}$ and estimate the scalar charges (\ref{eq:charges}) through a simple central finite difference approximation to the derivative operator:
\begin{align*}
\alpha_A^\textrm{approx} (\bar{m}_A) &= \frac{1}{m_A^{(0)}(\bar{m}_A)} \frac{m_A^{(+)}(\bar{m}_A)-m_A^{(-)}(\bar{m}_A)}{2\Delta \varphi},
\nonumber \\
\beta_A^\textrm{approx}  (\bar{m}_A) &= \frac{\alpha_A^{(+)}(\bar{m}_A)-\alpha_A^{(-)}(\bar{m}_A)}{2\Delta \varphi}, \\
k_A^\textrm{approx}(\bar{m}_A)  &= \frac{1}{I^{(0)}_A(\bar{m}_A)}\frac{I^{(+)}(\bar{m}_A) - I^{(-)}(\bar{m}_A)}{2\Delta \varphi}.
\nonumber \\
\end{align*}

\begin{figure}[t]
\includegraphics[width=0.9 \linewidth]{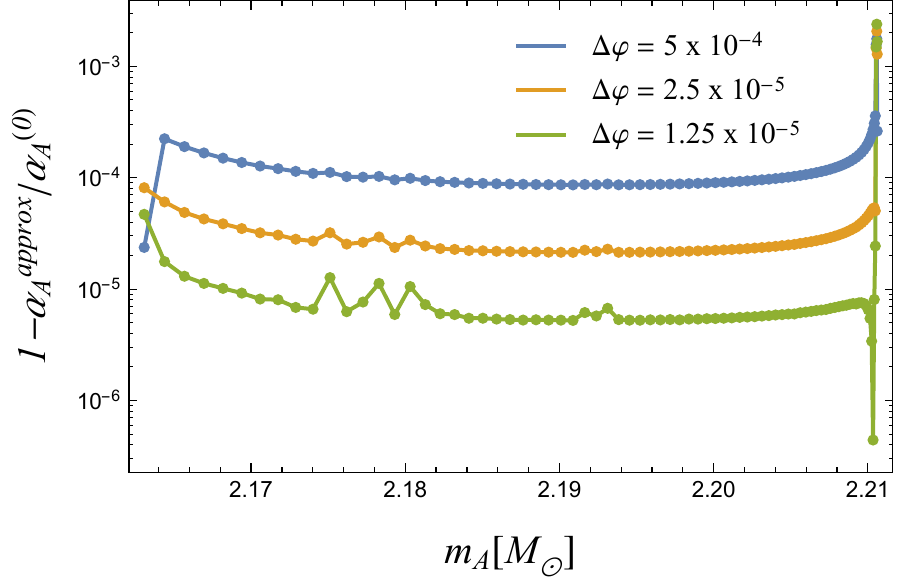}
\caption{Relative error $|(\alpha_A^{(0)} - \alpha_A^\textrm{approx})/\alpha_A^{(0)}|$ between the scalar charge computed  directly from the asymptotic behavior of the scalar field and the one computed using a finite difference approximation to the derivative operator. Here we employ the HC model with $\xi = 100$ and the ENG EoS, and consider the cases where $\Delta \varphi = 5 \times 10^{-4}$, $\Delta \varphi = 2.5 \times 10^{-4}$, and $\Delta \varphi = 1.25 \times 10^{-4}$. Using this data we can also determine the rate of convergence, which is found to agree with the second-order character of the finite difference stencil we use.}
\label{fig:conv1}
\end{figure}

Recall that the scalar charge $\alpha_A$ can either be computed from the expression above, or directly from the asymptotic behavior of the scalar field, through $\alpha_A = -\omega_A/m_A$. The latter procedure yields a much more reliable estimate for $\alpha_A$, and so the comparison between $\alpha_A^{(0)}$ and $\alpha_A^\textrm{approx}$ enables us to estimate the error incurred in the finite difference approximations above. This is shown in Fig.~\ref{fig:conv1}. For $\Delta \varphi = 5 \times 10^{-4}$, which is typically used in this work, we see that the relative error in the scalar charge is of the order of 0.01\%, but may increase near the boundaries of the mass interval where scalarized solutions exist. Errors of the same order are obtained for other values of $\xi$ as well.

Notice that, when presenting our results, we restrict consideration to the range of baryonic masses for which both $(+)$ and $(-)$ quantities exist. This may exclude from our consideration a narrow range of masses at the borders of the mass interval. These boundary values could be taken care of by a one-sided approximation to the derivative operator, which would lead, in particular, to a better resolution of the spikes in Fig.~\ref{fig:abk}. These fine details are not, however, too relevant for our analysis.

\bibliography{library}

\end{document}